\documentclass[11pt]{article}

\usepackage[utf8]{inputenc}
\usepackage[T1]{fontenc}
\usepackage[margin=1in]{geometry}
\usepackage[protrusion=true,expansion=false]{microtype}
\usepackage{authblk}
\usepackage{amsmath,amssymb}
\usepackage{graphicx}
\usepackage{booktabs}
\usepackage{tabularx}
\usepackage{array}
\usepackage{multirow}
\usepackage{longtable}
\usepackage{enumitem}
\usepackage{xcolor}
\usepackage{titlesec}
\usepackage[hidelinks,breaklinks=true]{hyperref}
\usepackage{url}

\setlist{nosep,leftmargin=*}
\titleformat*{\section}{\large\bfseries}
\titleformat*{\subsection}{\normalsize\bfseries}
\titleformat*{\subsubsection}{\normalsize\bfseries\itshape}

\title{\textbf{Authenticity Debt and the Synthetic Content Threat Landscape:} \\
A Layered Framework for Trust, Provenance, and IP Governance \\ in the Generative AI Era}

\author{
Shubhashis Sengupta\textsuperscript{1}, 
Benjamin McCarty\textsuperscript{1}, \\
Milind Savagaonkar\textsuperscript{1}, 
Rhine Andotra\textsuperscript{1} \\
\textsuperscript{1}Accenture Services Pvt. Ltd. \\
\texttt{
\{shubhashis.sengupta, benjamin.g.mccarty,\\
milind.savagaonkar, rhine.andotra\}@accenture.com
}
}

\date{}

\begin{document}
\maketitle

\begin{center}
\small
\textbf{arXiv classification:} \texttt{cs.CR} (primary); \texttt{cs.CY}, \texttt{cs.AI} (cross-list)
\end{center}
\vspace{0.5em}

\begin{abstract}
\noindent Generative artificial intelligence has fundamentally changed how content is now produced. It has enabled how high-fidelity text, images, audio, and videos are created, modified, and redistributed at near-zero marginal cost. This shift exposes enterprises and ecosystems to a number of risks across four reinforcing \textbf{authenticity layers}---authenticity, provenance, integrity, and accountability---that traditional controls are inadequate to address in isolation. We introduce the concept of \textbf{authenticity debt}: the cumulative institutional liability that accumulates when organizations deploy AI-generated content without preserving verifiable origin, integrity, and accountability, deferring exposure that surfaces under regulatory, legal, or market scrutiny. This paper presents a comprehensive, multi-dimensional taxonomy of generative AI harms and attack vectors, surveys the capabilities and failure modes of technical controls including digital watermarking, provenance frameworks (C2PA, Adobe CAI), and detection technologies, and argues that no single mechanism is sufficient in open, adversarial, and evolving environments. Drawing on Zero Trust Architecture principles and enterprise governance frameworks, we propose a layered reference architecture that integrates cryptographic provenance, human-in-the-loop verification, and continuous governance to sustain defensible authenticity at scale. We further examine the regulatory landscape (EU AI Act, U.S.\ FTC, NIST AI RMF) and identify practical guiding principles for organizations seeking to build authenticity as institutional infrastructure rather than an afterthought.
\end{abstract}

\vspace{1em}

\section{Introduction}

For decades, the digital economy relied on the fact that surface quality served as a practical proxy for authenticity. An email matching house style suggested legitimacy. A familiar voice implied identity. High production value conveyed authority. Generative AI has dismantled this assumption. Teams now produce high-fidelity text, audio, and video at near-zero marginal cost. Synthetic systems can replicate executive voices, draft regulatory documents in house style, and generate video that mirrors real events. \textbf{Appearance alone no longer serves as evidence of authorship.}

The practical consequences are already measurable. A finance executive joins what appears to be a routine video call with her CEO, who requests an urgent wire transfer. She approves it. Days later, she learns the CEO never made the request---an AI-generated voice and video had replicated him and bypassed internal controls. Variants of this scenario have cost organizations tens of millions of dollars globally. The U.S.\ Identity Theft Resource Center (ITRC) reported that impersonation scams surged 148\% from April 2024 to March 2025, while the Federal Trade Commission (FTC) recorded \$2.95 billion in reported losses to imposter scams in 2024 alone~\cite{itrc2025,ftc2025}.

These incidents reflect a structural shift in how authenticity functions. In the synthetic era, credibility is not a property of content; it is a property of systems. Enterprises that replace implicit trust with verifiable, architecturally embedded proof gain a durable advantage---what we term the \textbf{proof divide}---over those that continue to rely on surface cues.

The \textbf{Stanford HAI 2025 AI Index Report} highlights this acceleration: rapid improvements in generative model capabilities, falling inference costs, and a sharp rise in real-world AI-related incidents signal that misuse and unintended harms are scaling alongside adoption~\cite{stanford2025}. Controls designed for traditional digital media are systematically insufficient in an environment where content can be continuously transformed, regenerated, and recombined by AI systems.

This paper makes three primary contributions. \textbf{First}, we present a canonical, multi-dimensional taxonomy of generative AI harms and attacks that eliminates overlap across intent, attack vector, modality, attack technique, and affected trust layer. \textbf{Second}, we survey technical controls---watermarking, provenance standards, and detection technologies---with explicit attention to their capabilities, failure modes, and adversarial vulnerabilities. \textbf{Third}, we introduce and formalize the concept of authenticity debt and propose a layered reference architecture, grounded in Zero Trust principles, for building and sustaining enterprise authenticity infrastructure.

\paragraph{Organization.} Section~\ref{sec:landscape} characterizes the threat landscape. Section~\ref{sec:taxonomy} presents the taxonomy. Sections~\ref{sec:content-centric}--\ref{sec:trust-layers} analyze specific harm categories. Sections~\ref{sec:watermarking}--\ref{sec:detection} examine technical controls and their limits. Sections~\ref{sec:debt}--\ref{sec:architecture} develop the resilience architecture. Sections~\ref{sec:regulation}--\ref{sec:governance} cover regulation and governance. Sections~\ref{sec:scenarios}--\ref{sec:conclusion} present applied scenarios and conclusions.

\section{The New Threat Landscape in the Age of Generative AI}
\label{sec:landscape}

Generative AI is reshaping the intellectual property (IP) threat landscape by dramatically lowering the cost, speed, and scale at which content can be collected, transformed, and reproduced. Advances in foundation models, declining inference costs, and widespread deployment mean that high-quality text, images, audio, and video can now be generated or remixed at industrial scale. The implications extend far beyond isolated incidents: regulatory filings, intellectual property claims, investor communications, and internal approvals all depend on confidence in origin and integrity. As organizations deploy generative AI at scale, they increase both productivity and exposure simultaneously such that \textbf{content scales instantly while verification lags}.

\subsection{Mass Content Scraping and Transformation}
Large-scale web-scraped datasets form the backbone of many generative models, often without meaningful consent from original creators or publishers. This practice raises fundamental questions about ownership, permissible use, and downstream reuse of data embedded in model outputs~\cite{economictimes2024,brooklyn2024}. Investigations into AI training data markets suggest a growing ecosystem where data acquisition outpaces legal clarity.

\subsection{Erosion of Provenance and Authorship Norms}
Generative models can produce outputs that closely echo the style, structure, or expression of identifiable creators without attribution, making it difficult to determine origin, originality, or infringement~\cite{authorship2025}. AI systems can mimic distinctive creative styles and generate derivative works that challenge traditional copyright doctrines, which historically protect expression but not style per se~\cite{cc2023,trails2025}.

\subsection{The Widening Gap Between Practice and Principle}
While frameworks such as the OECD AI Principles emphasize transparency, accountability, and respect for intellectual property, \textbf{current generative AI systems often operate in ways that mask data lineage and weaken creator control}~\cite{oecd2024}. Shadow AI has now become a major concern for companies: employees using unsanctioned generative tools outside organizational visibility has quietly accelerated authenticity debt. Every AI-generated output produced outside governed workflows bypasses provenance, watermarking, and review controls, adding unverifiable content to the enterprise information environment with no audit trail and no accountability chain.

\section{A Multi-Dimensional Taxonomy of Generative AI Harms and Attacks}
\label{sec:taxonomy}

In many cases, AI-powered information attacks may compress or bypass stages of the traditional cyber kill chain~\cite{lockheed}. Because threat actors need only collect information to train or prompt AI systems to generate harmful outputs, the threat model is narrowed to adversary reconnaissance and delivery of weaponized information. We organize harms along five dimensions: intent, attack vector, modality, attack technique, and affected trust layer.

\begin{table}[!ht]
\centering
\small
\renewcommand{\arraystretch}{1.15}
\begin{tabularx}{\textwidth}{@{}l l X X@{}}
\toprule
\textbf{Dimension} & \textbf{Category} & \textbf{Description} & \textbf{Examples} \\
\midrule
\multirow{5}{*}{Intent}
 & Misinformation & Unintentional false or misleading content & Hallucinated citations, inaccurate summaries \\
 & Disinformation & Deliberate manipulation to mislead & Coordinated influence campaigns, Operation Doppelganger, Meliorator \\
 & Fraud \& Crime & Financial or identity-based exploitation & CEO voice fraud, synthetic KYC bypass \\
 & Impersonation \& Abuse & Identity misuse without direct financial gain & Fake journalists, brand impersonation \\
 & IP \& Data Misuse & Unauthorized use or leakage of protected content & Memorized training data, style cloning \\
\midrule
\multirow{4}{*}{Attack Vector}
 & Content-Centric & Harm from generated outputs & Deepfakes, fake news articles \\
 & Model-Centric & Harm from training, tuning, or reuse & IP leakage, data memorization, model poisoning \\
 & System-Centric & Exploits platforms or pipelines & Metadata stripping, C2PA manifest spoofing \\
 & Human-Centric & Exploits trust, cognition, or workflows & Social engineering, authority spoofing \\
\midrule
\multirow{5}{*}{Modality}
 & Text & Language-based generation & Fake press releases, phishing emails \\
 & Image & Synthetic or altered images & Face swaps, fabricated evidence \\
 & Voice & Synthetic or cloned speech & Vishing, executive impersonation \\
 & Video & Moving visual media & Fake speeches, staged events \\
 & Multimodal & Coordinated multi-modal attacks & Text + voice + video influence operations \\
\midrule
\multirow{4}{*}{Attack Technique}
 & Generation & Creating synthetic content & Model prompting, fine-tuning \\
 & Manipulation & Altering authentic content & Face reenactment, voice conversion \\
 & Amplification & Scaling reach and impact & Bot networks, automated posting \\
 & Evasion & Bypassing controls & Watermark removal, re-encoding, Diffusion Purification \\
\midrule
\multirow{4}{*}{Trust Layer}
 & Authenticity & Is the content real? & Deepfake media \\
 & Provenance & Where did it come from? & Metadata stripping, private key exfiltration \\
 & Integrity & Was it altered? & Partial manipulation, archive tampering \\
 & Accountability & Who is responsible? & Anonymous or laundered content, spoofed signatures \\
\bottomrule
\end{tabularx}
\caption{Multi-Dimensional Taxonomy of Generative AI Harms and Attacks.}
\label{tab:taxonomy}
\end{table}

\section{Content-Centric Threats: Misinformation, Disinformation, and Deepfakes}
\label{sec:content-centric}

Content-centric threats arise from the generation and distribution of synthetic or manipulated media. Nation-state and cybercriminal actors have operationalized these capabilities at scale. The Storm-1516 group created AI deepfakes disseminated as ``leaked'' tapes to interfere with electoral processes~\cite{microsoft2024}. The Spamouflage group leveraged large language models to generate inauthentic comments in coordinated information operations~\cite{openai2025}. Emerald Sleet (Kimsuky) used AI to generate deepfake military identification documents and spearphishing emails~\cite{kimsuky2024}. The Doppelganger group mimicked news media and produced AI-generated articles on look-alike domains~\cite{doj2024}. CopyCop employed a private Llama-3 instance fine-tuned on politically biased data to rewrite authentic news at scale while simultaneously attempting to poison other LLM training corpora~\cite{recordedfuture2024}. Cybercriminal actors have also marketed malicious LLM-based tools built exclusively for offensive use---such as FraudGPT, sold on dark-web marketplaces and Telegram---to enable phishing, social engineering, and fraud at scale~\cite{fraudgpt}.

\textbf{These incidents share a common structure}: AI dramatically reduces the cost and expertise required to produce convincing disinformation, lowers the bar for targeted impersonation, and enables operations to scale across multiple communication channels simultaneously. The attack surface is therefore not bound by the sophistication of individual actors but by the accessibility of generative infrastructure.

\subsection{Reconnaissance and Delivery}

Adversaries typically begin by collecting data to enable deepfakes, voice cloning, writing style impersonation, and target-specific knowledge. Delivery occurs across email, websites, USB drives, social media, phone calls, and video calls. The effectiveness of these attacks is fundamentally dependent on the target consuming and accepting the information as authentic---cognitive manipulation is the final attack surface.

\subsubsection{Reconnaissance: Gathering Target Information}

Adversaries collect high-quality media and behavioral data on intended targets to train or prompt AI systems. The goal is to accumulate sufficient signal to enable convincing deepfakes, voice cloning, writing style impersonation, and exploitation of tacit knowledge that increases attack plausibility. For high-value targets (HVTs) such as executives, public officials, or prominent journalists, this data is often already publicly available through prior media appearances, conference recordings, and social media, making exposure-limitation impractical in most cases. Three reconnaissance-phase mitigations merit consideration, each with meaningful operational constraints:

\begin{itemize}
  \item \textbf{Exposure Limitation.} Attempt to limit exposure of high-quality media and data on likely HVTs by reducing unnecessary public-facing audio, video, and image content. This mitigation aligns with the MITRE D3FEND\texttrademark{} Isolate Group technique~\cite{d3fend-isolate} but is generally impractical for individuals with existing substantial public presence. Modern deepfake tools can construct convincing impersonations from a few seconds of audio or a single photograph, meaning the threshold for sufficient training data is already met for most public-facing executives.
  \item \textbf{False Persona Deception Campaigns.} Introduce carefully crafted false personas as deception bait to lure threat actors into targeting HVTs that do not exist (aligned with the MITRE D3FEND\texttrademark{} Deceive Group technique~\cite{d3fend-deceive}). This approach requires active maintenance to construct and sustain a persona convincing enough that adversaries invest resources in targeting it. If successful, it consumes adversary reconnaissance capacity without exposing genuine principals. The primary limitation is operational cost: maintaining a credible false identity at the fidelity required to attract sophisticated actors is resource-intensive and may be infeasible outside of intelligence-community or high-security contexts.
  \item \textbf{Data Poisoning.} Poison legitimate images and data with hidden adversarial perturbations designed to disrupt a threat actor's ability to train a convincing deepfake model on the target. Research by the Australian Federal Police and Monash University has demonstrated this approach in practice~\cite{afp-monash}. The principal limitation is reversibility: motivated adversaries can often detect deepfake failures caused by poisoned training data and, in many cases, remove or compensate for the perturbation, particularly as adversarial robustness techniques improve. Poisoning should therefore be treated as a friction-raising measure rather than a reliable barrier.
\end{itemize}

\subsubsection{Delivery: Transmitting Weaponized Information}

Once the AI-generated or AI-augmented attack payload has been constructed, adversaries deliver or disseminate it through communication channels selected to maximize target exposure and minimize detection. Delivery vectors include email, websites, USB drives, social media platforms, phone calls, and video calls. The effectiveness of the attack is contingent not on the technical sophistication of the payload alone but on the target's willingness to consume and accept the information as authentic. This final cognitive layer is the ultimate attack surface, and mitigations at the delivery phase therefore span both technical controls and human behavioral interventions. Five delivery-phase mitigations merit structured consideration:

\begin{itemize}
  \item \textbf{Threat Intelligence and Domain Monitoring.} Utilize threat intelligence capabilities to monitor posts, news articles, accounts, mentions, domain names, and other open-source publicly observable indicators to detect delivery of weaponized information from unofficial accounts, enabling takedown requests and counter-messaging (aligned with the MITRE D3FEND\texttrademark{} Domain Registration Takedown technique~\cite{d3fend-domain}). The effectiveness of this mitigation is bounded by the breadth and currency of intelligence collection and the speed of countermeasure execution against the threat actor's operational tempo.
  \item \textbf{User Awareness and Security Training.} Include advanced AI-powered attacks and harm scenarios in user awareness and security training programs to reduce the cognitive vulnerability that makes delivery effective. Training should emphasize recognition of urgency and authority manipulation patterns, critical evaluation of unexpected or out-of-band requests, and escalation procedures for suspicious communications. This mitigation addresses the cognitive attack surface directly and is necessary regardless of technical control quality, since even well-defended systems rely on humans making correct trust decisions at the margin.
  \item \textbf{Strong Authentication Resistant to AI-Based Attacks.} Implement phishing-resistant multi-factor authentication (MFA) with hardware-level security (e.g., Trusted Platform Modules), cryptographic identity verification, and biometric authentication controls that are resistant to replay and impersonation attacks (aligned with MITRE D3FEND\texttrademark{} Agent Authentication and Credential Hardening techniques~\cite{d3fend-credential}). Authentication mechanisms should be evaluated specifically for resistance to AI-generated voice or video impersonation, as traditional knowledge-based or behavioral challenges are increasingly defeatable by advanced generative models.
  \item \textbf{Message Integrity and Cryptographic Verification.} Utilize security measures to ensure the integrity of information is not compromised or tampered with during delivery (aligned with MITRE D3FEND\texttrademark{} Message Hardening~\cite{d3fend-message}). This mitigation requires published authoritative records---for example, email SPF and DKIM records for message authentication, or C2PA cryptographically protected signatures for media content---or decentralized identifiers that allow observers to independently verify the integrity and authenticity of received information against trusted authoritative sources. Without such mechanisms, delivered content carries no independently verifiable claim of origin.
  \item \textbf{Curated High-Credibility Information Feeds.} Curate and filter automated information consumption feeds including news aggregators and AI training data pipelines to source exclusively from high-credibility publishers that themselves implement authoritative security controls. Preferred sources should support cryptographic integrity protection, provenance verification, and traceability back to original sources. This is particularly important for organizations where AI systems consume external information feeds that could be weaponized through CopyCop-style contamination strategies, where adversary-generated disinformation is deliberately injected into sources intended to train or inform downstream AI systems.
\end{itemize}

Taken together, these delivery-phase mitigations are not mutually exclusive and are most effective when layered. \textbf{No single countermeasure is sufficient}: technical integrity controls can be bypassed by compromised signers; authentication controls can be defeated by advanced impersonation; training reduces but cannot eliminate cognitive vulnerability. The adversary's ultimate objective is to induce the target to act on false information as though it were true---mitigations succeed when they impose sufficient friction, uncertainty, or independent verification requirements to interrupt that chain before irreversible action is taken.

\section{Model-Centric Risks: Data Misuse, Memorization, and IP Leakage}
\label{sec:model-centric}

Unlike user-driven misuse, model-centric risks are embedded in the training and operational design of generative AI systems. They manifest across five primary mechanisms: morphing and editing of copyrighted creative elements~\cite{morphing2024}, attribution stripping during dataset preprocessing~\cite{ldm2024}, paraphrasing and regeneration of protected text~\cite{poisonedparrot}, and synthetic misattribution of AI-generated content~\cite{misattribution2024}, automated scraping without consent.

Memorization represents a particularly significant risk: large language models can reproduce verbatim or near-verbatim fragments of training data under certain prompting conditions, potentially exposing proprietary, sensitive, or third-party-owned content in outputs delivered to end users~\cite{carlini2021}. Fine-tuning and transfer learning amplify this risk by re-exposing proprietary data in new model deployments.

Collectively, these behaviors illustrate that IP risk is not confined to user actions but is structurally embedded in how generative models are trained and deployed. Addressing them requires careful dataset curation, memorization mitigation, robust data governance, and enforceable IP frameworks that extend to the model level.

\section{Why Authenticity and Provenance Matter: Four Layers of Enterprise Trust}
\label{sec:trust-layers}

Authenticity in the synthetic era is not a binary property of individual content items. It is a systemic property that rests on four reinforcing and mutually dependent elements:

\begin{itemize}
  \item \textbf{Authenticity:} whether content genuinely originates from the claimed source.
  \item \textbf{Provenance:} the ability to trace content back to its origin across transformations and platforms.
  \item \textbf{Integrity:} whether content has changed since creation, and whether those changes are documented.
  \item \textbf{Accountability:} who authorized, approved, or published the content, and under what governance.
\end{itemize}

A weakness in any single layer reduces overall confidence. Weaknesses across multiple layers compound exposure nonlinearly. Authenticity debt---defined as the cumulative institutional liability created when organizations deploy AI-generated content without preserving verifiable origin, integrity, and accountability---builds silently at the intersection of these gaps~\cite{itrc2025,economictimes2024}.

C2PA cryptographic protections can provide strong guarantees against tampering: if a manifest is signed and the signature validates, the content is cryptographically proven unmodified since signing. However, C2PA cannot protect against compromised credential issuers, prevent signers from signing inauthentic data, or defend against sufficiently advanced cryptanalytic attacks. Unsigned content is unverified---it is not proven false, but neither is it proven authentic. A content-agnostic trust model must therefore integrate cryptographic verification with governance controls over who may sign and under what circumstances.

\section{Digital Watermarking: Capabilities, Trade-offs, and Limits}
\label{sec:watermarking}

Digital watermarking embeds information within content to signal origin, authenticity, or ownership. In the context of generative AI, watermarking has emerged as a proactive mechanism to identify AI-generated content across modalities. Its effectiveness is constrained by robustness limits, performance trade-offs, and adversarial vulnerability. This section evaluates techniques across modalities and situates watermarking within the broader trust stack.

\subsection{Text Watermarking}
Two broad categories of text watermarking exist. LLM-integrated watermarking modifies the token selection process during generation so that the output sequence carries a statistically detectable pattern without degrading semantic quality, a prominent example being Google's SynthID Text~\cite{synthid}. Pre- and post-generation methods embed patterns through whitespace modulation, Unicode manipulation, keyword substitution, or semantic watermarking~\cite{wm-survey,wm-impossibility}. \textbf{Text watermarking} faces severe trade-offs: stronger signals risk altering readability; stealthy patterns may be too weak for reliable detection; and post-generation attacks including paraphrasing, translation, and model-based rewrites significantly reduce detectability~\cite{wm-survey}.

\subsection{Image Watermarking}
Image watermarking techniques span spatial domain methods (Least Significant Bit, Patchwork), frequency-domain methods (DCT, DFT, DWT, SVD), and hybrid combinations that maximize robustness against adversarial attacks~\cite{image-wm}. Frequency-domain methods generally survive JPEG compression, Gaussian noise, and mild filtering. All techniques exhibit sensitivity to geometric transformations, AI-based regeneration, and adversarial evasion. Stronger watermarks improve detectability but may introduce perceptible artifacts; hybrid methods improve robustness at the cost of computational complexity.

\subsection{Audio Watermarking}
Audio watermarking employs spread-spectrum methods, phase or echo embedding, and psychoacoustic masking to hide signals imperceptible to listeners~\cite{audio-wm}. Spread-spectrum and psychoacoustic methods survive MP3 compression, background noise, and mild filtering, but are vulnerable to re-recording, pitch changes, time-stretching, and AI-based audio regeneration.

\subsection{Video Watermarking}
Video watermarking extends image techniques across temporal dimensions using frame-based embedding, temporal distribution across multiple frames, and hybrid spatial-temporal methods~\cite{video-wm}. Temporal and hybrid methods survive compression and basic editing but are vulnerable to cropping, frame removal, re-encoding, and AI-based video enhancement. Computational overhead is significantly higher than for image or audio watermarking.

\subsection{Soft-Binding: Integrating Watermarks with C2PA}
A key architectural insight is that C2PA provenance and digital watermarks can be combined in a soft-binding relationship. A C2PA manifest hash can be embedded as a watermark within the content itself. This watermarked data can then be used to query against a C2PA manifest API to recover the original manifest and compare a computed fingerprint against the watermarked data. This approach is resistant to transfer attacks that attempt to move one watermark onto another file to evade C2PA validation. In this configuration: signed, valid, and soft-bound content is verified authentic; signed and valid but unbound content confirms manifest integrity without content-level binding; unsigned but soft-bound content with valid watermark can still recover provenance metadata through the API channel.

\section{Provenance Frameworks and Standards}
\label{sec:provenance}

\subsection{C2PA (Coalition for Content Provenance and Authenticity)}
C2PA is an open standard for capturing, signing, and verifying provenance metadata for digital content. It embeds metadata in a tamper-evident manifest cryptographically linked to content, supporting assertions about creator identity, toolchain, creation and modification timestamps, and AI involvement. Verification confirms authenticity only if the signature matches a trusted signing key. Trust can break if metadata is removed, altered, or if a tool in the chain does not follow the standard~\cite{c2pa-spec}.

\subsection{Adobe Content Authenticity Initiative (CAI)}
CAI operationalizes the C2PA specification within Adobe tools and workflows. Metadata is collected automatically during creation or modification, signed by the creator's key, and shared with content~\cite{adobe-aca}. Third parties can verify signatures without accessing internal workflow details. The initiative strengthens provenance claims for AI-generated content through tamper-evident metadata binding integrated with widely used creative software~\cite{adobe-cai}.

\subsection{Emerging Standards}
Additional standards and initiatives include W3C Verifiable Credentials (focused on credential claims in a decentralized trust model), Digimarc and Truepic (proprietary watermarking and proof-of-origin solutions), the MPAI Neural Network Watermarking Standard (interoperability and evaluation metrics for AI-specific watermarking), and the ITU AI and Multimedia Authenticity Collaboration (cross-industry international guidance). None of these have achieved the adoption level of C2PA.

\subsection{Trust Limitations}
Provenance metadata represents a claim, not a guarantee. Without verification, any tool can falsify creator or tool information. Digital signatures enable trust, but that trust is contingent on verified signing keys, toolchain adherence, and governance practices. If any tool in the workflow removes or modifies metadata, the chain of trust breaks. Proper key management, identity verification, and workflow integration are necessary conditions for reliable provenance.

\section{Attack and Evasion Techniques}
\label{sec:evasion}

Digital watermarking bypass techniques range from simple operations---cropping, pixel averaging, lossy compression, file format conversion, analog-to-digital transfer, translation-based text reconstruction---to sophisticated adversarial attacks. Diffusion-based image editing has been demonstrated to defeat robust invisible watermarks through the ``Diffusion Washing'' technique, which re-imagines content through an improved Latent Diffusion Model, removing noise-based signals including Google's SynthID~\cite{diffusion-washing}. Research has further shown that robust watermarks can leak extractable information, enabling adversarial watermark manipulation through channel-aware feature extraction~\cite{wm-leak}.

Several sophisticated attacks have also been identified or theorized against C2PA systems. The Nikon Overlay attack (sometimes called the ``Washing'' attack) attempts to overwrite C2PA manifests. The AWS Bedrock Titan ``Mask'' attack targets private key infrastructure. Spoofed C2PA manifests and certificates---exemplified by the Leica Camera AG case---represent a class of credential impersonation attacks. Theorized attacks include high-resolution screen replay into a C2PA-secured camera and Adversary-in-the-Middle (AitM) injection of a deepfake signal into camera hardware at the circuit level. The ``Maskirovka'' technique---drawing on Russian military deception doctrine---involves creating easily detectable AI-generated signals to lure detectors into complacency, then deploying previously unrevealed stealth techniques on strategic operations.

A fundamental asymmetry governs this domain: defenders must secure against all attack vectors simultaneously, while attackers need only find one successful bypass. Detection tools are often available to attackers who can continuously test and mutate their content until it evades detection. Digital data composed of binary sequences carries no intrinsic proof of origin or authenticity, which threat actors exploit to produce arbitrary binary states indistinguishable from authentic ones.

\section{Detection Technologies: Capabilities and Failure Modes}
\label{sec:detection}

Detection technologies aim to identify synthetic or manipulated content after creation and dissemination. Unlike provenance and watermarking, which are origin-based controls, detection is reactive and probabilistic---inherently fragile in adversarial environments. Detection functions best as a supporting signal within a broader trust architecture, not as a primary trust mechanism.

\subsection{Detection Modalities}
Visual detection tools identify signs of digital alteration or artificial generation through visual inconsistencies, unnatural motion, or generation artifacts. NIST Media Forensics and FaceForensics++ benchmarks demonstrate effective performance in controlled settings, with reliability declining as content is edited, compressed, or shared across platforms~\cite{faceforensics}. Voice deepfake detection focuses on anomalies in tone, rhythm, and speaker identity; ASVspoof benchmarks show accuracy declining sharply with short clips, background noise, or newer voice generation technologies~\cite{asvspoof}. Text detection tools estimate AI generation probability from writing patterns and statistical signals, producing high false-positive rates that become unreliable after light editing, translation, or human-AI combination~\cite{text-detection-limitations}.

\subsection{Systemic Failure Modes}
Across modalities, detection systems exhibit consistent weaknesses: poor generalization to unseen models and techniques; performance degradation after compression, resizing, or post-processing; susceptibility to adversarial evasion; and limited explainability that undermines legal defensibility. Public evaluations by NIST and DARPA Semantic Forensics show that detection accuracy drops sharply outside controlled conditions~\cite{nist-omf,darpa-semafor}. Detection performs better when content is created and analyzed within the same platform or ecosystem, where proprietary signals are available---advantages that are lost once content crosses platform boundaries.

\section{Authenticity Debt: Formalizing the Accumulation of Trust Risk}
\label{sec:debt}

As generative AI systems scale, failures in authenticity and attribution controls do not remain isolated. Unresolved limitations in watermarking, provenance, and detection accumulate over time, creating authenticity debt: the deferred institutional risk associated with relying on trust signals that are known to be incomplete, fragile, or context-dependent~\cite{itrc2025,economictimes2024}.

The concept is formally analogous to technical debt in software engineering~\cite{tech-debt}: just as deferred resolution of known system limitations leads to cumulative risk and escalating remediation cost, deferred resolution of authenticity weaknesses leads to increasing downstream costs for investigation, dispute resolution, compliance, and legal defensibility. Authenticity debt grows as organizations continue to deploy AI outputs, models, and workflows without revisiting underlying trust assumptions, as content is edited and redistributed, as attribution signals are lost or contested across platforms, and as shadow AI bypasses governed workflows entirely.

Four organizational profiles illustrate the debt accumulation pattern: financial institutions reviewing AI-generated documents lacking traceability to source models; healthcare and pharmaceutical firms assessing training datasets with contested provenance; legal teams evaluating client-facing outputs potentially containing memorized third-party proprietary content; and advisory firms unable to document which model generated a deliverable, what training data informed it, or who authorized its release.

\textbf{Regulatory pressure reinforces the urgency}. EU AI Act Article 53 requires detailed training data summaries from general-purpose AI model providers; Article 50 mandates machine-readable labeling of synthetic media~\cite{eu-ai-act}. These obligations create enforceable accountability anchored directly to an organization's ability to demonstrate provenance and integrity. They are currently in force.

Managing authenticity debt requires treating trust as a lifecycle risk rather than a one-time control problem. Organizations must periodically reassess trust assumptions, limit upstream IP exposure, and rely on authoritative records and governance processes to anchor accountability as technical controls degrade or fail over time.

\section{Reference Architecture for Trust and Authenticity Systems}
\label{sec:architecture}

AI-driven misinformation, disinformation, and malinformation attacks fundamentally target trust. They aim to deceive individuals or systems into accepting inauthentic content as genuine. Traditional mitigation strategies relying on AI-content detectors or fraud detection strategies are likely to become increasingly unreliable against sophisticated attacks as generative AI capabilities continue to advance. A more \textbf{robust approach adapts the principles of Zero Trust Architecture (ZTA)}~\cite{nist-zta}, combined with strong governance processes, to systematically eliminate implicit trust and enforce continuous verification of information and identity.

\subsection{Zero Trust Core Principles}
Consistent with NIST SP 800-207~\cite{nist-zta} and the DoD Zero Trust Reference Architecture, \textbf{systems should operate under the assumption of breach}: organizations perpetually function within hostile environments where adversaries may have compromised users, systems, or data flows. This requires treating all information, requests, and communications as potentially fabricated; verifying authenticity, integrity, and authorization explicitly for every transaction regardless of source; and assuming that infrastructure, data stores, user accounts, or communication channels may already be under adversary control.

\subsection{Four Pillars of Resilient Enterprise Architecture}
\begin{itemize}
  \item \textbf{Assume Control Failure.} Attribution mechanisms including watermarking, provenance metadata, and detection are treated as probabilistic signals rather than guarantees. The organization does not rely on any single mechanism as the sole basis for IP or authenticity decisions.
  \item \textbf{Limit IP Exposure Upstream.} Access to proprietary data, models, and high-risk generation capabilities is governed through least-privilege controls, segmentation, and logging. By constraining who can train, fine-tune, or extract outputs, the organization reduces the likelihood and impact of IP leakage before content leaves trusted environments.
  \item \textbf{Preserve Defensible Records.} The organization maintains authoritative, cryptographically protected records documenting data rights, model lineage, and generation or access events. These records serve as the primary evidentiary anchor for audits, enforcement actions, and ownership disputes, even when downstream content has been stripped of attribution signals.
  \item \textbf{Enable Investigation and Response.} Systems and processes are designed for forensic readiness, including retention of relevant logs, defined escalation paths, and coordination between technical, legal, and operational teams. This ensures the organization can investigate suspected IP misuse, preserve evidence, and respond in a timely and defensible manner.
\end{itemize}

\subsection{Authoritative Publication Channels}
A core architectural recommendation for mitigating MDM (Misinformation, Disinformation, and Malinformation) threats is the establishment of secure, verifiable official communication channels that serve as a single source of truth, deterring disinformation by making it easy for stakeholders to verify authentic information and negate false information as inconsistent or unverifiable. These channels should implement cryptographically signed metadata (origin, timestamp, hashes, author) directly embedded in content, support self-referencing hashes enabling third-party verification even when content is copied or shared, and apply Zero Trust segmentation to prevent unauthorized access by insider threats or advanced persistent threats.

\subsection{Human-in-the-Loop Controls}
High-risk or high-privilege actions should require explicit policy enforcement through \textbf{Policy Decision Points} acting as failsafe. Key governance controls include mandatory sanity checks before proceeding, dual or multi-party authorization for sensitive actions, deliberate human-in-the-loop steps that break reliance on potentially compromised digital channels, and Out-of-Band (OOB) authentication that confirms identity through a separate secure channel before acting on requests received via email, phone, video call, or messaging. Multi-Source Cross-Verification using multiple independent access paths protected by different security mechanisms reduces the feasibility of simultaneous AitM attacks across all sources~\cite{us-patent}.

\section{Regulatory and Policy Landscape}
\label{sec:regulation}

The regulatory environment governing synthetic content is evolving rapidly across jurisdictions, with enforceable obligations already in force in the European Union and emerging frameworks in the United States.

The \textbf{EU AI Act Article 50} mandates that providers of AI systems generating synthetic media must mark outputs in machine-readable format to ensure detection as artificially generated. \textbf{Article 53} requires providers of general-purpose AI models to publish detailed summaries of training data. The \textbf{EU Copyright in the Digital Single Market Directive (CDSM) Article 3} permits scientific research organizations to mine data for AI training without licenses, while \textbf{Article 4} prohibits training on content where rights holders have expressly reserved their rights~\cite{eu-cdsm}.

In the United States, \textbf{NIST AI 100-4} provides voluntary guidance on reducing risks posed by synthetic content through technical approaches to digital content transparency~\cite{nist-ai-100-4}. The FTC has finalized rules banning AI deepfake impersonation of government and business entities. The \textbf{TAKE IT DOWN Act} criminalizes the creation and distribution of nonconsensual intimate imagery, including deepfakes, with a 48-hour removal mandate upon notice~\cite{take-it-down}. These developments represent an accelerating trend toward enforceable accountability anchored to provenance and integrity claims.

\section{Governance, Risk, Accountability, and Human Vigilance}
\label{sec:governance}

Effective mitigation of AI-related harms requires clear ownership, enforceable accountability, and sustained human vigilance. Technical controls alone are insufficient; organizations must operationalize governance structures that align decision rights, escalation paths, and human oversight with the real-world failure modes of AI systems. The sections below address each governance dimension in turn, drawing on NIST, ISO, OECD, and regulatory frameworks as normative anchors.

\subsection{Ownership and Accountability}
Organizations should assign explicit ownership for AI systems across their full lifecycle, encompassing data sourcing, model development, deployment, and ongoing use. \textbf{Accountability should be tied to existing enterprise roles with real decision authority rather than delegated to advisory bodies alone}. The NIST AI Risk Management Framework (Govern function) calls for clear role definition, decision authority, and accountability for AI risks throughout the system lifecycle~\cite{nist-airmf}. ISO/IEC 23894 on AI Risk Management emphasizes responsibility assignment for AI risk identification, mitigation, and monitoring~\cite{iso-23894}. The EU AI Act imposes documented responsibility and oversight obligations specifically for high-risk AI systems~\cite{eu-ai-act}.

Critically, \textbf{ownership must include authority to act}: responsible parties should have the standing to pause deployments, escalate incidents, and initiate remediation when harms---such as IP misuse, impersonation, or misinformation campaigns---are suspected or detected. Ownership without commensurate authority creates accountability gaps that adversaries and regulators alike can exploit.

\subsection{Escalation Paths and Incident Handling}
AI-related harms often emerge gradually and ambiguously rather than as discrete detectable events. Organizations should \textbf{define pre-established escalation paths that connect frontline detection to legal, compliance, security, and executive decision-makers} before incidents occur. NIST SP 800-61 provides recommendations for incident response, including escalation, containment, and recovery, that are directly applicable to AI misuse and IP incidents~\cite{nist-800-61}. The NIST AI RMF (Manage function) further emphasizes documented processes for responding to AI risk events and integrating them into existing enterprise risk and incident response workflows~\cite{nist-airmf}.

A key design principle for escalation thresholds is that they should be based on risk indicators, not certainty. Because attribution and detection signals for AI-generated content are probabilistic rather than definitive, waiting for conclusive evidence before escalating systematically delays response and increases harm. Escalation criteria should be calibrated to the cost asymmetry between a false positive (unnecessary review) and a false negative (unmitigated harm), deviating toward earlier escalation for high-impact decision domains.

\subsection{Human Oversight and Review}
\textbf{Human oversight is required for high-impact AI decisions}, particularly where legal exposure, reputational risk, or IP ownership is implicated. The OECD Trustworthy AI Principles require meaningful human oversight for AI systems affecting rights, trust, or accountability~\cite{oecd2024}. The NIST AI RMF (Measure and Manage functions) recommends human-in-the-loop or human-on-the-loop controls where automated systems operate under uncertainty or in high-impact domains~\cite{nist-airmf}.

In practice, \textbf{human oversight should be focused on exceptions, escalations, and dispute resolution rather than routine automation}. Attempting to apply human review to every AI-generated output is operationally infeasible at scale and creates a false sense of coverage. The governance design challenge is therefore to identify the specific decision points where human judgment adds irreplaceable value---high-stakes approvals, contested provenance claims, external communications with legal or regulatory significance---and to concentrate oversight capacity there.

\subsection{Training and Vigilance Programs}
Organizations should implement role-based training programs ensuring employees understand AI risks, control limitations, and reporting procedures. The NIST AI RMF (Govern function) identifies workforce education and awareness as necessary enablers of effective AI governance~\cite{nist-airmf}. The OECD AI Principles similarly emphasize organizational capability and awareness as core components of responsible AI deployment~\cite{oecd2024}. \textbf{Training should not be treated as a one-time compliance exercise: it must be continuous and updated as AI capabilities, threat actor techniques, and organizational deployment patterns evolve}.

Program design should differentiate by role: finance and accounts-payable staff require deepfake fraud scenario training; communications and public affairs teams require disinformation and impersonation awareness; technical and data teams require IP and memorization risk training; and executive and board-level stakeholders require strategic framing of authenticity debt and regulatory exposure. Periodic simulation exercises---including tabletop scenarios involving AI-generated executive impersonation or synthetic disinformation events---are particularly effective at building institutional muscle memory for escalation and response.

\subsection{Operational Guardrails and Use Policies}
Documented policies should define acceptable and prohibited uses of AI systems, with particular specificity around high-risk capabilities such as synthetic voice and video generation, automated external communications, and training data sourcing. The NIST AI RMF recommends governance policies covering data use, model outputs, and deployment contexts~\cite{nist-airmf}. The FTC's own AI compliance plan illustrates such internal governance in practice: a data and AI governance board sets AI use policies and assesses risks, non-public data may not be exposed to generative AI tools that could train on it, and non-compliant tools can be terminated~\cite{ftc-ai}.

Guardrails must be enforceable through both technical and procedural controls. \textbf{Policy statements alone are insufficient}: effective guardrails combine access controls that prevent unauthorized use at the system level, logging and monitoring that detect policy deviations, and procedural checkpoints that require explicit authorization before high-risk capabilities are invoked. Shadow AI---use of unsanctioned generative tools outside organizational visibility---represents a direct policy enforcement gap that requires both detection capability and clear consequence frameworks to address.

\subsection{Review, Metrics, and Continuous Improvement}
AI governance should be reviewed regularly to ensure controls remain effective as models, data, threat actor techniques, and organizational usage patterns evolve. The COSO Enterprise Risk Management framework treats risk management as a continuous process supported by monitoring, feedback, and corrective action rather than a periodic audit cycle~\cite{coso-erm}. ISO 31000 similarly stresses ongoing review of controls in response to changing risk conditions~\cite{iso-31000}. Neither framework is AI-specific, but both provide the continuous improvement discipline that AI governance requires given the pace of capability and threat evolution.

Organizations should track a small, defensible set of governance metrics rather than attempting comprehensive instrumentation. \textbf{Recommended metrics include: control coverage rate} (percentage of AI systems subject to defined governance controls and policy enforcement); \textbf{escalation timeliness} (time from initial risk signal to formal escalation or decision); \textbf{policy exception frequency} (number and trend of approved and unapproved deviations from AI use policies); \textbf{training completion and refresh rate} (percentage of relevant staff completing required AI risk training on schedule); \textbf{incident and near-miss volume} (reported AI-related incidents and near-misses over time, disaggregated by harm category); and \textbf{remediation cycle time} (time to implement corrective actions following identified control gaps). These metrics are deliberately oriented toward process quality and response effectiveness rather than technical detection rates, which are subject to the evasion dynamics described in Section~\ref{sec:detection} and should not be treated as primary governance indicators.

\section{Guiding Principles and Applied Threat Scenarios}
\label{sec:scenarios}

\subsection{AI Deepfake Executive Impersonation}
When an executive appears to request a large fund transfer, sharing of sensitive information, or urgent action via phone, video call, or messaging, the following mitigations are effective against advanced deepfakes. \textbf{Out-of-Band verification is the primary control}: contact the executive using a pre-established trusted channel (company-issued secure phone with phishing-resistant MFA), treat any urgent, unscheduled, or abnormal request as suspicious until verified, and maintain an integrity-protected list of signatories and payees with a direct-confirmation change-control policy. \textbf{Pre-agreed secret safe words} known only to a small, trusted group provide an additional but non-sufficient authentication factor.

Organizations should avoid over-relying on the following countermeasures, which may provide false assurance against highly motivated adversaries: limiting high-quality public audio/video of executives (modern deepfake tools work from seconds of audio or a single photo), poisoning audio/video data (attackers can detect and remove many poisoning techniques), real-time liveness tests (easily defeated by advanced AI models), digital watermarks or C2PA alone (weaponizable if a trusted device is compromised), and multi-person approval processes (advanced attackers may simultaneously deepfake multiple executives in the same call).

\subsection{AI-Enhanced Disinformation, Misinformation, and Malinformation}
MDM (Misinformation, Disinformation, and Malinformation) threats using AI to generate, amplify, or disseminate harmful information are becoming larger, more sophisticated, and more precisely tailored to evade detection and exploit cognitive biases. The most effective mitigation is establishing a secure, verifiable official communication channel as a single source of truth. This channel should implement cryptographically protected publication workflows, digitally sign all official content with immutable protections, embed cryptographically signed metadata directly into content, and apply Zero Trust segmentation to prevent unauthorized access. Enterprise-wide communication policies should mandate that all official information be cross-vetted for accuracy and published only through the protected channel, with organizational prohibitions on making official statements outside that channel.

\section{Conclusion}
\label{sec:conclusion}

Generative AI fundamentally alters the scale, speed, and ambiguity of intellectual property and authenticity risk. Synthetic content can be created, modified, and redistributed at marginal cost, while traditional signals of authorship, ownership, and authenticity degrade rapidly across platforms and over time. No single technical control such as watermarking, provenance metadata, or detection can reliably establish origin or ownership in open, adversarial, and evolving environments.

Sustaining trust in the AI ecosystem requires a \textbf{layered approach in which technical mechanisms serve as probabilistic signals within a broader system of controls}. Organizations must assume control failure, limit upstream IP exposure, preserve authoritative records, and design systems for investigation and response. Governance, accountability, and human vigilance are not optional complements to technical controls; they are essential for maintaining defensible positions when attribution is contested or controls break down.

The concept of \textbf{authenticity debt provides a unifying framework}: unresolved trust gaps accumulate over time, increasing operational friction, legal exposure, and long-term risk. Left unmanaged, IP leakage and contested ownership can undermine competitive advantage, weaken legal defensibility, trigger regulatory and contractual exposure, and erode stakeholder trust at scale. As synthetic content volume increases, scrutiny will \textbf{intensify}: regulators will formalize traceability expectations, courts will demand defensible evidence of origin, and clients will expect verification in high-impact interactions. The institutions that will lead are those that \textbf{design for proof}---embedding verification into system architecture rather than relying on after-the-fact reconstruction.

Ultimately, the challenge of trust and IP protection in the age of generative AI is not only technical. It is organizational and systemic. Institutions that combine realistic assumptions about control limits with disciplined governance, clear ownership, durable records, and informed human judgment will be better positioned to manage disputes, sustain credibility, and responsibly scale AI capabilities in a contested information environment.


\end{document}